\documentclass[pre]{revtex4}
\usepackage{amsmath}
\usepackage{graphicx}
\begin{document}
\title{Numerical Study of Crystal Size Distribution in Polynuclear Growth}
\author{Hidetsugu Sakaguchi and Takuma Ohishi}
\affiliation{Department of Applied Science for Electronics and Materials,
Interdisciplinary Graduate School of Engineering Sciences, Kyushu
University, Kasuga, Fukuoka 816-8580, Japan}
\begin{abstract}
The crystal size distribution in polynuclear growth is numerically studied using a coupled map lattice model. The width of the size distribution depends on $c/D$, where $c$ is the growth rate at interface sites and $D$ is the diffusion constant. When $c/D$ is sufficiently small, the width $W$ increases linearly with $c/D$ and saturates at large $c/D$. Monodisperse square and cubic crystals are obtained respectively on square and cubic lattices when $c/D$ is sufficiently small for a small kinetic parameter $b$. The linear dependence of $W$ on $c/D$ in a parameter range of small $c/D$ is explained by the eigenfunction for the first eigenvalue in a two-dimensional model and a mean-field model. For the mean-field model, the slope of the linear dependence is evaluated theoretically. 
\end{abstract}
\maketitle
\section{Introduction}

Crystal growth has been extensively studied in various research fields such as applied physics, metallurgy, chemical engineering, and mineralogy. It is also studied in statistical physics as a typical nonequilibrium phenomenon. 
Nucleation, surface kinetics, and diffusion are important factors in crystal growth~\cite{rf:1}. Crystal growth is one of the Stefan problems for moving interfaces, which is difficult to analyze mathematically.  
Various simulation methods have been proposed to study crystal growth far from equilibrium. The phase-field model is a partial-differential equation in which the solid and melt phases are expressed with a continuous variable corresponding to the order parameter~\cite{rf:2,rf:3}. We proposed a coupled map lattice model for crystal growth~\cite{rf:4,rf:5,rf:6}.  A similar type of order parameter is defined on a lattice, and coupled maps are used for the time evolution. The numerical simulation of the coupled map lattice is much faster than that using the phase-field model. One reason is that  the width of the interface between the solid and melt phases is only one lattice constant in the coupled map lattice model. On the other hand, the order parameter changes smoothly at the interface in the phase-field model owing to the diffusion term in the equation for the order parameter.  We performed numerical simulation of crystal growth of dendrites and diffusion-limited aggregations (DLA's) using coupled map lattices. These complicated fractal patterns can be easily reproduced in the coupled map lattice model, because the surface tension effect can be set to zero in the coupled map lattice model. In previous papers, we studied crystal growth starting from a single crystal seed. In this paper, we study polynuclear growth in which there are many crystal seeds initially, and investigate the crystal size distribution at the final stationary state. 

Crystal size distribution (CSD) is one of important topics in crystallography. Various theories of CSD were proposed. Becker and D\"oring studied the time evolution of the size distribution in the nucleation process~\cite{rf:7}. Randolph and Larson discussed stationary CSDs in the research field of chemical engineering~\cite{rf:8}. Marsh applied the CSD theory in geology and found an exponential size distribution of mineral crystals deposited from magma~\cite{rf:9}. In their model, the exponential distribution is derived by the balance of the outflow, the inflow, and the growth of crystals. 
In the synthesis of micro- and nanocrystals, the final size distribution after the crystal growth is important. Various methods of characterization and control  of the crystal size distribution  have been studied. In particular, the control of the crystal size distribution have been investigated in colloidal sciences~\cite{rf:10,rf:11}. The size distributions similar to the gamma distribution or the log normal distribution are often observed in experiments. The size distribution $P(S)$ is not symmetric around the average value but has a longer tail for large $S$.  
It is practically important to produce almost monodisperse crystals, because the monodisperse crystals are useful for various applications such as catalysts, magnetic fluids, and drug delivery systems. Colloidal crystals can be produced from monodisperse microcolloidal particles, which can be applied to various systems such as photonic crystals~\cite{rf:12}. 

The crystal size distribution depends on many factors such as nucleation processes, surface conditions of crystals, and various experimental conditions. The CSD problem is still an open problem. For the synthesis of monodisperse crystals, it is important that the time scales of nucleation and growth processes are separated, and the size distribution of the crystal seeds is almost mono-disperse at the initial stage of the growth process. In this paper, we consider a problem that the size distribution of the crystal seeds is almost monodisperse at the initial stage of the crystal growth, but the initial positions of the crystal seeds are randomly distributed. The final size distribution changes owing to the competition of the diffusion and surface kinetics. Our model might be applied to a two-dimensional crystal growth on a substrate in that the positions of crystals are fixed in time. However, the purpose of this study is to understand qualitatively a condition wherein the final crystal size distribution becomes almost monodisperse.      
\section{Two-Dimensional Coupled Map Lattice Models for Polynuclear Growth}
We first study a two-dimensional coupled map lattice model on a square lattice.
There are two variables in our coupled map lattice for the solution growth, namely, the order parameter $x_n(i,j)$ and the concentration $u_n(i,j)$, where  $(i,j)$ and $n$ denote respectively a lattice site and a discrete time. 
The order parameter $x_n(i,j)$ is 1 at crystal sites, 0 at solution sites, and between 0 and 1 at interface sites. 
The time evolution has two stages, namely, diffusion and surface kinetics. The diffusion is expressed as
\begin{equation}
u_n^{\prime}(i,j)=u_n(i,j)+D\{u_n(i+1,j)+u_n(i-1,j)+u_n(i,j+1)+u_n(i,j-1)-4u_n(i,j)\},
\end{equation}
where $D$ is a diffusion constant. The surface kinetics is expressed as
\begin{eqnarray}
x_{n+1}(i,j)&=&x_n(i,j)+cbu^{\prime}_n(i,j),\nonumber\\
u_{n+1}(i,j)&=&u^{\prime}_n(i,j)-cbu^{\prime}_n(i,j),
\end{eqnarray}
where $c$ is a parameter for the growth rate. The time evolution Eq.~(2) is applied only at interface sites satisfying $0<x_n(i,j)<1$. 
If $x_n(i,j)$ goes over a critical value 1, interface sites change into crystal sites and the neighboring solute sites are assigned to interface sites. By repeating Eqs.~(1) and (2), polynuclear crystal growth occurs. 
The parameter $b$ is  1 at nonflat sites and takes a small value $b_0$ at flat sites.  The flatness of an interface site is determined by counting the number of crystal sites in the four nearest-neighbor sites of the interface site. If the number is 1, $b$ is set to be $b_0$, and if the number is larger than 2,  $b$ is set to be 1. For sufficiently small $b_0$, square crystals appear, because kink sites and concave sites are quickly occupied, but it takes a rather long time for flat sites to grow. The parameter $b$ represents an effect of the surface tension on the surface kinetics. In the case of $b_0=1$, the surface tension effect disappears. In Eq.~(2), the increase in the order parameter $x$ is equal to the absorption of $u$ at interface sites, that is, the conservation law of $x+u$ is satisfied. In our model of Eqs.~(1) and (2), only the deposition process from solution to crystal is assumed, that is, the inverse process of dissolution is not taken into consideration.

\begin{figure}[t]
\begin{center}
\includegraphics[height=7.cm]{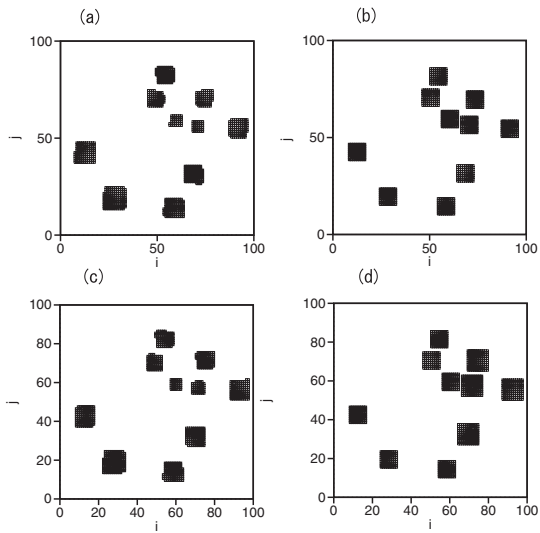}
\end{center}
\caption{Crystal sites  at (a) $D=0.01$ and $c=1$, and (b) $D=0.2$ and $c=0.01$ in the case that the initial sizes of crystals are all 1. Crystal sites  at (c) $D=0.01$ and $c=1$, and (d) $D=0.2$ and $c=0.01$ in the case that the initial sizes of crystals are 1 for $i<70$ and 9 for $i>70$.}
\label{f1}
\end{figure}
\begin{figure}[t]
\begin{center}
\includegraphics[height=3.7cm]{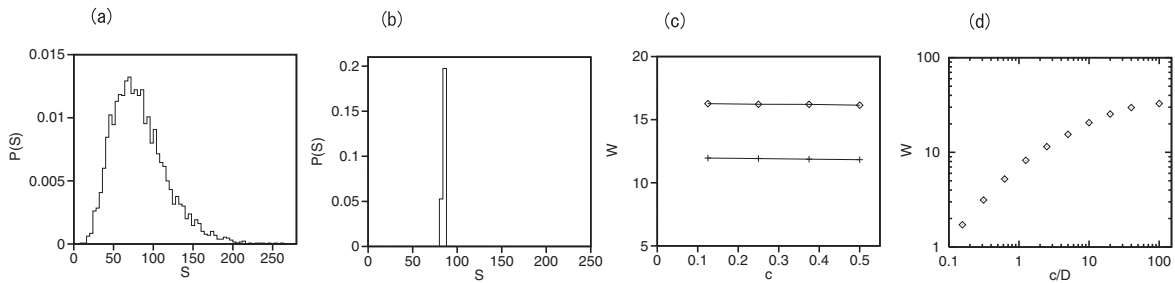}
\end{center}
\caption{Size distributions at (a) $D=0.01$ and $c=1$, and (b) $D=0.2$ and $c=0.01$. (c) Relationship between $W$ and $c$ for $c/D=5$ (rhombi) and $c/D=2.5$ (pluses). (d) Relationship between $W$ and $c/D$.}
\label{f2}
\end{figure}

We have performed numerical simulation of the two-dimensional coupled map lattices.  Initially, 10 crystal seeds of size 1 are randomly set in a square of $L\times L=100\times 100$. The initial value of $u(i,j)$ is uniform and set to be $u_0=0.1$, and the parameter $b$ is assumed to be 0.005.  Periodic boundary conditions are imposed. After a long-time iteration of Eqs.~(1) and (2), $u$ becomes zero owing to the diffusion and the absorption of $u$ at interface sites, and the crystal growth process is completed. 
Figure 1(a) shows crystal sites at $D=0.01$ and $c=1$ at the final time. The crystal size is widely distributed.  The crystal sizes are small in a region where initial seeds are densely distributed.  In our model, the randomness originates only from the initial configuration of crystal seeds. 
Figure 1(b) shows crystal sites at $D=0.2$ and $c=0.01$. Crystals take a clear square shape, and the sizes are almost uniform. As $D$ is large and $c$ is small, the concentration $u$ tends to be uniform and the crystal sizes become homogeneous even for the random initial configuration of seeds. 
To show an effect of the initial seed size distribution, the sizes of four crystal seeds at $i>70$ are set to be $3\times 3=9$, and the sizes of the other seeds are set to be 1.   Figures 1(c) and 1(d) show crystal sites at (c) $D=0.01$ and $c=1$ and (d) $D=0.2$ and $c=0.01$ at a final time. At $D=0.2$ and $c=0.01$, the crystal sizes for the six seeds of size 1 are the same as those in Fig.~1(b), and the crystal sizes for the four larger seeds of size 9 are larger by $11^2-9^2=40$ than those in Fig.~1(b).  At $D=0.01$ and $c=1$, the crystal sizes change for all crystals from the values in Fig.~1(a), but the change in size from Fig.~1(a) depends on the position. The final crystal size depends on the initial seeds, but the dependence of the final size distribution on the initial size distribution is not so simple. We investigate only the case that the initial sizes are all 1 hereafter. 

We have calculated the CSD using a larger square lattice of $400\times 400$.  The size $S$ is defined as the sum of $x_n(i,j)$ in crystal sites and interface sites around each crystal seed. At crystal sites, $x_n(i,j)$ takes a value of 1. Then, $S$ represents the area of a two-dimensional crystal. Figures 2(a) and 2(b) show the size distribution at (a) $D=0.01$ and $c=1$, and (b) $D=0.2$ and $c=0.01$. (There are a few crystals that collide with neighboring crystals, but the sizes of those crystals are not considered in the size distribution. The collision probability decreases if the system size increases for a fixed number of crystal seeds. In the production of colloidal particles, the electric double layer around colloidal particles prevents the collision.) The size distribution is rather wide at $D=0.01$ and $c=1$. The distribution is asymmetric and has a longer tail for large $S$.  The size distribution is very narrow at $D=0.2$ and $c=0.01$, which implies that crystals are almost monodisperse.  

The average of crystal size is estimated as $u_0L^2/N$, because the total sum of $S$ is equal to $L^2u_0$ owing to the conservation law of $u+x$. 
The inhomogeneity of the crystal size is characterized by the width $W$ or the standard deviation of the size distribution. Figure 2(c) shows the widths $W$ for four different $c$'s at the same $c/D=2.5$ and $5$. The width $W$ depends strongly on $c/D$ but hardly depends on $c$ under the condition of $c/D=$const.  
If $c$ and $D$ are small and the one-step increments of $x$ and $u$ in Eqs.~(1) and (2) are sufficiently small, Eqs.~(1) and (2)  can be approximated by differential equations. It can be shown that the final stationary states depend only on the ratio $c/D$ owing to a scale transformation of time if the differential approximation is good.  Figure 2(d) shows a relationship between $W$ and the ratio $c/D$ in a logarithmic plot. The diffusion constant $D$ is changed as $0.01,0.025,0.05,0.1$ and $0.2$ for $c=1$, and then $c$ is changed as $1/2^n$ ($n=1,2,\cdots, 6$) for $D=0.2$. The width increases linearly as $W\sim 10.6(c/D)$ in a range of small $c/D$ and tends to saturate at large $c/D$. That is, monodisperse crystals are obtained in the limit of $c/D=0$, and the dispersion is proportional to $c/D$ when $c/D$ is sufficiently small.  

\begin{figure}[t]
\begin{center}
\includegraphics[height=3.7cm]{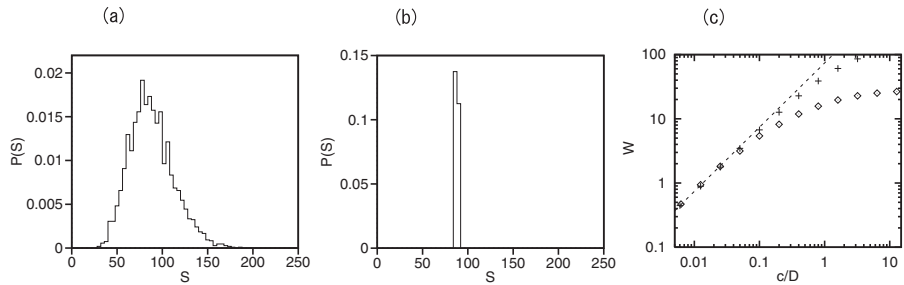}
\end{center}
\caption{Size distributions at (a) $D=0.025$ and $c=0.64$, and (b) $D=0.2$ and $c=0.00125$ for Eqs.~(1),(4), and (5). (c) Relationship between $W$ and $c/D$.}
\label{f3}
\end{figure}

In the coupled map lattice model of Eqs.~(1) and (2), there is a possibility that neighboring crystals become closer to each other and collide with each other as they grow. To neglect the effect of the collision, we can propose a simpler coupled map lattice model.  Equation (2) is rewritten as
\begin{eqnarray}
x_{n+1}(i,j)&=&x_n(i,j)+c^{\prime}u^{\prime}_n(i,j)/r_{n}(i,j)\times 2\pi r_n(i,j),\nonumber\\
u_{n+1}(i,j)&=&u_{n+1/2}(i,j)-c^{\prime}u^{\prime}_n(i,j)/r_n(i,j)\times 2\pi r_n(i,j),
\end{eqnarray}
where $ c^{\prime}=cb/(2\pi)$, and $x_n(i,j)$ and $r_{n}(i,j)=\sqrt{x_n(i,j)/\pi}$ are interpreted respectively as the area and radius of circular crystals. Here, the growth velocity of crystals of radius $r_n(i,j)$ is assumed to be inversely proportional to the radius $r_n(i,j)$ as $v_r=c^{\prime}/r$. This type of relation is satisfied under the condition of diffusion-limited growth. Equation (3) is expressed as 
\begin{eqnarray}
x_{n+1}(i,j)&=&x_n(i,j)+cu^{\prime}_n(i,j),\\
u_{n+1}(i,j)&=&u^{\prime}_n(i,j)-cu^{\prime}_n(i,j),
\end{eqnarray}
where $2\pi c^{\prime}$ is rewritten as $c$. 
The form of Eqs.~(4) and (5) is almost equal to that of Eq.~(2); however, the time evolution of Eqs.~(4) and (5) is applied only at the positions of crystal seeds in this model. That is, in contrast to Eq.~(2), the variable $x$ increases indefinitely even if $x$ becomes larger than 1. Growth patterns such as square patterns do not appear, because $x$ grows only at the seed points. The magnitude of $x$ is interpreted as crystal size. In other words, we consider a situation that the radii of crystals are sufficiently small in contrast to the distances among different crystals, and the growth patterns are invisible. 
In the following numerical simulation, the number $N$ of initial seeds is set to be 180, and the system size is $L\times L=400\times 400$.  The crystal growth stops for sufficiently large $n$, when $u_n(i,j)$ decreases to 0.  Figure 3(a) shows the crystal size distribution at $D=0.025$ and $c=0.64$. Figure 3(b) shows the size distribution at $D=0.2$ and $c=0.00125$. When $c$ is small and $D$ is large, the size distribution becomes narrow. The rhombi in Fig.~3(c) show the relationship between the width $W$ of the size distribution and the ratio $c/D$, where $c$ is changed as $5.12/2^n$ ($n=1,2,\cdots, 12$) for $D=0.2$. The dashed line denotes $74(c/D)$. The width $W$ increases from 0 linearly in a parameter region of small $c/D$ and tends to saturate at large $c/D$ also in this simplified model.

Because Eqs.~(1) and (5) are linear equations, an initial value problem for $u$ can be solved using the eigenvalues and eigenfunctions as $u_n(i,j)=\sum_k a_ke_k(i,j)(\lambda_k)^n$, where $\lambda_k$ and $e_k(i,j)$ are eigenvalues and eigenfunctions, and $a_k$ is the expansion coefficient. The first eigenvalue $\lambda_1$ and the corresponding eigenfunction $e_1(i,j)$ can be numerically evaluated by a long-time iteration of the same equation Eqs.~(1) and (5) and the normalization $u_{n+1}(i,j)\rightarrow u_{n+1}(i,j)/\lambda_1$. When $c/D$ is sufficiently small, the first eigenvalue is close to 1 and the first component of $a_1e_1(i,j)(\lambda_1)^n$ decays slowly or the other components $\sum_{k\ne 1}a_ke_k(i,j)(\lambda_k)^n$ decay quickly. Under this condition, the growth rate of each crystal is expected to be proportional to the first eigenfunction $e_1(i,j)$ at the crystal-seed point, because the concentration $u$ decays as $e_1(i,j)\lambda_1^n$ and $x(i,j)$ increases to compensate for the decrease in $u$. The crystal-size distribution can be evaluated approximately using numerically estimated $e_1(i,j)$ by assuming that the crystal size is proportional to the eigenfunction $e_1(i,j)$ and  using the fact that the average  crystal size is equal to $L^2u_0/N$. The pluses in Fig.~3(c) show the width of the size distribution evaluated from $e_1(i,j)$. The width is close to the results denoted by rhombi obtained by direct numerical simulation of Eqs.~(1), (4), and (5) for small values of $c/D$. At large $c/D$, the contribution from the other eigenfunctions cannot be neglected, and some deviation is observed. 

\section{Voronoi Tessellation and a Mean-Field Model of Polynuclear Growth}
The Voronoi tesselation is a method of space partitioning. In the construction  of the Voronoi tessellation, $N$ central points are randomly distributed, and the whole plane is partitioned into territories of each point. The boundary line between two territories is determined by the perpendicular bisector of the two central points.  Each Voronoi cell is a set of points surrounded by  perpendicular bisectors. The Voronoi tesselation is applied to various areas such as grains of crystals and territories of animals. The size distribution of Voronoi cells is approximately given by the gamma distribution:~\cite{rf:13,rf:14} 
\begin{equation}
P(S_v)\propto S_v^{\gamma}e^{-\alpha S_v}, 
\end{equation}
where $S_v$ is the area of each Voronoi cell, $\alpha$ is a parameter, and $\gamma\sim 2.5$.  

In our model, the crystal grows fast by absorbing the surrounding solute when $c/D$ is sufficiently large. The absorption range or the territory of each crystal seed might be approximated by the Voronoi cell.  We interpret that the central point in each Voronoi cell corresponds to a seed point of a crystal. 
First, we consider a case that all the solutes in each Voronoi cell are absorbed into the central seed of a crystal. 
In that case, the crystal size $S$ is proportional to the area $S_v$ of the Voronoi cell. In that case, the crystal-size distribution is expressed as  
\begin{equation}
P(S)\propto S^{\gamma}e^{-\alpha S}. 
\end{equation}
 The parameter $\alpha$ is determined from the average value of $S$.
In our model, the average size of crystals is given by $u_0 L^2/N$, because of the conservation of the total mass of solute. 
At $\gamma=2.5$, the average value can be calculated as 
\begin{equation}
\langle S\rangle = \frac{\int_0^{\infty}S^{\gamma+1}e^{-\alpha S}dS}{\int_0^{\infty}S^{\gamma}e^{-\alpha S}dS}=\frac{1}{\alpha}\frac{\Gamma(\gamma+1)}{\Gamma(\gamma)}=\frac{7}{2\alpha},
\end{equation}
from the definition of the gamma function. The parameter $\alpha$ is given by 
\begin{equation}
\alpha=\frac{7N}{2u_0L^2}\sim 0.0393
\end{equation}
for $N=180,u_0=0.1$, and $L=400$. 
Figure 4(a) shows the distribution of Eq.~(7) for $N=180,u_0=0.1$, and $L=400$. The size distributions in Figs.~2(a) and 3(a) are qualitatively close to the size distribution shown in Fig.~4(a), in that there is a longer tail in a region of $S>\langle S\rangle$.  The average value in the size distribution in Fig.~3(a) is close to that in Fig.~4(a) because the number of initial crystal seeds is the same; however,  the width of the size distribution in Fig.~3(a) is smaller than that of the size distribution of the Voronoi tessellation.   

\begin{figure}[t]
\begin{center}
\includegraphics[height=3.7cm]{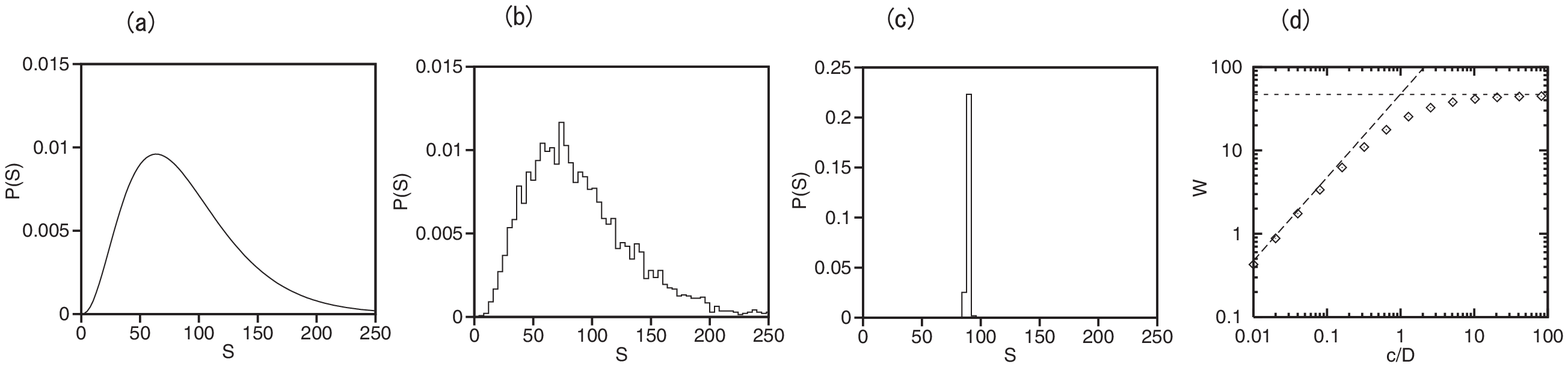}
\end{center}
\caption{(a) Size distribution determined by the Voronoi tessellation where $\alpha$ is given by Eq.~(9).  (b) Size distribution determined using Eqs.~(1), (10), and (11) at $D=0.01$ and $c=1$.  (c) Size distribution determined using the mean-field model at $D=1$ and $c=0.02$. (d) Relationship between the width $W$ and $c/D$.}
\label{f4}
\end{figure}

If $c/D$ is not so large, the effect of diffusion needs to be taken into consideration. To understand qualitatively the dependence of the width of the size distribution on $c/D$ using an even simpler model, we propose a mean-field type model for crystal growth, keeping the Voronoi tessellation in mind. The model is coupled differential equations expressed as 
\begin{eqnarray}
\frac{dx_i}{dt}&=&cu_i,\\
\frac{du_i}{dt}&=&\frac{1}{S_{vi}}\{-cu_i+D(\bar{u}-u_i)\},
\end{eqnarray}
where $x_i$ denotes the crystal size in the $i$th Voronoi cell, $u_i$ is the concentration of the solute in the $i$th Voronoi cell, $\bar{u}=(1/N)\sum u_i$, and $S_{vi}$ denotes the size of the $i$th Voronoi cell. The crystal is assumed to grow only at the seed point located at the center of the $i$th Voronoi cell with the growth rate $cu_i$, which is similar to the previous model using Eqs.~(1), (4), and (5). 
The concentration inside each Voronoi cell is assumed to be uniform and denoted as $u_i$. The concentrations $u_i$ in different Voronoi cells are different but becomes uniform owing to the mean-field-type diffusion term $D(\bar{u}-u_i)$ in Eq.~(11). The total mass $\sum_i (x_i+u_iS_{vi})$ is conserved in the time evolution of Eqs.~(10) and (11). In our numerical simulation, $S_{vi}$ is determined by the Voronoi tessellation of 180 seeds in a $400\times 400$ square lattice. The crystal size $S_i$ is defined as $x_i$ at the final stationary state of 
Eqs.~(10) and (11). Figure 4(b) is the size distribution of $S_i$ at the final time at $c=1$ and $D=0.01$. A rather wide size distribution close to the distribution shown in Fig.~4(a) is obtained at this parameter set satisfying $c/D=100$. Figure 4(c) is the size distribution of $S_i$ at $c=0.02$ and $D=1$. The size distribution is very narrow, when $c/D$ is sufficiently small. Figure 4(d) shows the relationship between the width $W$ of the size distribution and $c/D$. Similarly to previous models, we find the linear dependence of $W$ on $c/D$ in the small range of $c/D$ and the saturation at large $c/D$. 

The concentration $u_i$ can be expanded using eigenvalues $\lambda_k$ ($k=1,2,\cdots,N)$ and the corresponding eigenfunctions $e_{k,i}$ of the linear equation (11) as $u_i(t)=\sum_{k=1}^N a_ke_{k,i}e^{-\lambda_k t}$ where $a_k$'s are the expansion parameters. Substitution of this expression into Eq.~(11) yields
\begin{equation}
e_{k,i}=\frac{D\sum_{j=1}^{N} e_{k,j}}{N(c-\lambda_kS_{vi}+D)}.
\end{equation}
By the summation of Eq.~(12) from $i=1$ to $N$, the eigenvalue $\lambda$ satisfies
\begin{equation}
\frac{1}{N}\sum_{j=1}^{N}\frac{D}{c+D-\lambda S_{vj}}=1.
\end{equation}
This is the $N$th-order algebraic equation and there are $N$ solutions corresponding to the $N$ eigenvalues. 
If $c=0$, $\lambda=0$ is a solution to Eq.~(13).
For sufficiently small $c$ and small $\lambda$, Eq.~(13) is rewritten as
\[1-\frac{c}{D}+\frac{\lambda}{D}\frac{1}{N}\sum_{j=1}^{N}S_{vj}=1,\]
using the Taylor expansion. The first eigenvalue $\lambda$ is therefore expressed as 
\begin{equation}
\lambda=\frac{c}{\bar{S_v}},
\end{equation}
where $\bar{S_v}=(1/N)\sum_{j=1}^{N} S_{vj}$. 
If $c$ is sufficiently small, the first eigenvalue $\lambda_1$ is small, and  
 $u_{i}$ can be approximated at $u_i=a_1e_{1,i}e^{-\lambda_1t}$ because the other terms decay quickly. Owing to Eq.~(12), $e_{1,i}$ is proportional to $1/(c-\lambda_{1}S_{vi}+D)$, and $u_i(t)$ is expressed as 
\begin{equation}
u_{i}(t)=\frac{u_{i0}}{c+D- cS_{vi}/\bar{S_v}}e^{-\lambda_1 t}.
\end{equation}
Since $u(0)=u_0$, $u_{i0}=u_0(c+D-cS_{vi}/\bar{S_v})\sim u_0D$ for a sufficiently small $c$.  
The final stationary value of $x_i$ is given by the time integration of $cu_i(t)$ as
\begin{equation}
S_i=\int_0^{\infty}\frac{cu_{i0}}{c+D- cS_{vi}/\bar{S_v}}e^{-\lambda_1 t}\sim\frac{\bar{S_v}u_0D}{c+D-cS_{vi}/\bar{S_v}}\sim \bar{S_v}u_0\left (1+\frac{c}{D}\frac{S_{vi}-\bar{S_v}}{\bar{S_v}}\right ).
\end{equation}
The average value of $S_i$ is $\bar{S_v}u_0$ and the width $W$ of the size distribution of $S_i$ is given by 
\begin{equation}
W=\frac{c}{D}\delta S_v u_0,
\end{equation}
where $\delta S_v$ is the root mean square of the distribution of $S_{vi}$. The width $W$ is evaluated as $W=47.5 (c/D)$, since $\delta S_v u_0$ is evaluated as $\sqrt{2/7}(L^2/N_0) u_0=47.5$ using a formula of the gamma function.  The theoretical result is shown by a dashed line in Fig.~4(d). Fairly good agreement is seen in a region of sufficiently small $c/D$. We have shown again that the eigenfunction for the largest eigenvalue determines the width of the size distribution for sufficiently small $c/D$. For sufficiently large $c/D$,  the total solute $u_0S_{vi}$ in the $i$th Voronoi cell is absorbed to the $i$th seed at the center. This is the special situation corresponding to Fig.~4(a). The crystal size is evaluated as $S_{vi}u_0$. Then, the width of the size distribution is expected to be $W=\delta S u_0=47.5$, which is plotted by the dotted line in Fig.~4(d). 

In this simple model, $x_i$ increases in proportion to $u_i$ and the dynamics of $u_i$ is determined only by $\{u_j\}$. Therefore, if the initial sizes $x_i(0)$ are not zero but randomly distributed, the final crystal sizes $x_{if}$ are calculated as $x_{if}=x_{if0}+x_i(0)$, where $x_{if0}$ is the final crystal size in the case of $x_i(0)=0$.  That is, the final size distribution is determined by the convolution of the initial size distribution of $x_i(0)$ and the distribution of $x_{if0}$. 

\section{Three-dimensional model}
In this section, we study a three-dimensional coupled map lattice, which is more realistic than the two-dimensional model. 
The three-dimensional extension of Eq.~(1) on a cubic lattice is expressed as

\begin{eqnarray}
u_n^{\prime}(i,j,k)&=&u_n(i,j,k)+D\{u_n(i+1,j,k)+u_n(i-1,j,k)+u_n(i,j+1,k)+u_n(i,j-1,k)\nonumber\\
& &+u_n(i,j,k+1)+u_n(i,j,k-1)-6u_n(i,j,k)\}.
\end{eqnarray}
\begin{figure}[t]
\begin{center}
\includegraphics[height=3.7cm]{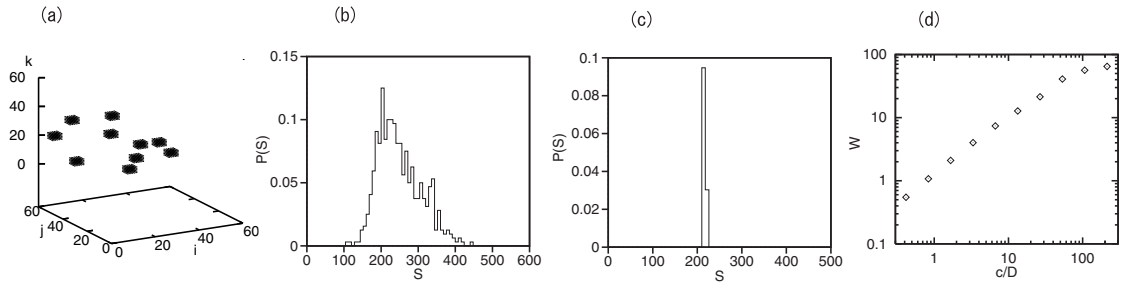}
\end{center}
\caption{(a) 3D plot of the crystalized site at $D=0.15$ and $c=0.1$ for Eqs.~(18) and (19).  (b) Size distribution at $D=0.01875$ and $c=1$. (c) Size distribution at $D=0.15$ and $c=0.125$. (d) Relationship between the width $W$ and $c/D$ for $c=1$.}
\label{f5}
\end{figure}
The three-dimensional extension of Eq.~(2) is expressed as
\begin{eqnarray}
x_{n+1}(i,j,k)&=&x_n(i,j,k)+cbu^{\prime}_n(i,j,k),\nonumber\\
u_{n+1}(i,j,k)&=&u^{\prime}_n(i,j,k)-cbu^{\prime}_n(i,j,k),
\end{eqnarray}
where $(i,j,k)$ denotes a lattice point in a cubic lattice, and the parameter $b$ is  1 for nonflat sites and takes a small value $b_0=0.005$ for flat sites. Here, the flat interface site is a site that has only one crystal site and five solution sites as its nearest-neighbor sites.  The system size is set to be $60\times 60\times 60$.  
 Figure 5(a) shows a 3D plot of the crystal site obtained in the numerical simulation of the three-dimensional coupled map lattice at $D=0.15$ and $c=0.1$. 
Initially, 11 crystal seed points are randomly distributed.
Clear cubic crystals are created, and the sizes of crystals are almost uniform. 
The size distributions are calculated for 8 samples of initial 100 seeds of crystals. Figure 5(b) shows  the size distribution at $D=0.01875$ and $c=1$. Figure 5(c) shows  the size distribution at $D=0.15$ and $c=0.125$. Similarly to the two-dimensional models, the size distribution becomes wider as $c/D$ increases. 
Figure 5(d) shows the relationship between the width $W$ of the size distribution and $c/D$. Here, $D$ is changed as $D=0.3/2^n$ ($n=1,2,\cdots,5$) for $c=1$, and then $c$ is changed as $1/2^n$ ($n=1,2,\cdots,5$) for $D=0.15$. The width increases linearly with $c/D$, which is similar to the results found in the two-dimensional models.

\section{Summary}
We have numerically studied the crystal size distribution in polynuclear growth using coupled map lattice models. The size distribution of crystals originates from the initial random configuration of crystal seeds, and the dispersion is reduced by the effect of the diffusion. We have found that the width of the size distribution is approximately determined by the parameter $c/D$, where $c$ is the growth rate at the interface sites and $D$ is the diffusion constant. When $c/D$ is sufficiently small, the width $W$ increases linearly with $c/D$ and tends to saturate at large $c/D$. The numerical simulations were performed on a square lattice and a cubic lattice. Monodisperse square and cubic crystals are obtained when $c/D$ are sufficiently small and the kinetic parameter $b$ is sufficiently small. The research and development of the manufacturing process of monodisperse crystals, especially monodisperse nanocrystals,  are important for various applications.  The linear dependence of $W$ on $c/D$ in the parameter range of small $c/D$ is explained by the eigenfunction for the first eigenvalue in a simplified two-dimensional model and an even more simplified mean-field-type model. For the mean-field model using the Voronoi tessellation, the slope of the linear dependence has been evaluated theoretically.

\end{document}